\documentclass[11pt]{aastex}
\usepackage{emulateapj5,apjfonts}
\usepackage{onecolfloat}
\usepackage{epsf}

\shortauthors{Bell 2002}
\shorttitle{Exponential Disks in LSB Galaxies}

\newcommand{\ha}{{\rm H$\alpha$ }}
\newcommand{\hi}{{\rm H{\sc i} }}
\newcommand{\hans}{{\rm H$\alpha$}}

\newcommand{\msun}{${\rm M_{\sun}}$}

\slugcomment{{\sc Accepted by the Astrophysical Journal : } 14th August 2002 }

\begin{document}


\def\head{

\title{Exponential Stellar Disks in Low Surface Brightness Galaxies:
   A Critical Test of Viscous Evolution} 

\author{Eric F. Bell}
\affil{Steward Observatory, University of Arizona, 933 N. Cherry Avenue, 
   Tucson, AZ 85721, USA}
\email{ebell@as.arizona.edu}

\begin{abstract}
\vspace{0.2cm}
Viscous redistribution of mass in Milky Way-type galactic disks 
is an appealing way of generating an exponential stellar profile
over many scale lengths, almost independent of initial conditions, 
requiring only that the viscous timescale and star
formation timescale are approximately equal.  However, galaxies
with solid-body rotation curves cannot undergo viscous
evolution.  Low surface brightness (LSB) galaxies have exponential surface
brightness profiles, yet have slowly rising, nearly
solid-body, rotation curves.  Because of this, 
viscous evolution may be inefficient in LSB galaxies: the
exponential profiles, instead would give important insight into
initial conditions for galaxy disk formation.

Using star formation laws from the literature, and tuning the
efficiency of viscous processes to reproduce an exponential stellar
profile in Milky Way-type galaxies, I test the role of viscous
evolution in LSB galaxies.  Under the conservative and not
unreasonable condition that LSB galaxies are gravitationally
unstable for at least a part of their lives, I find that it
is impossible to rule out a significant r\^ole
for viscous evolution.  This type of model
still offers an attractive way of producing exponential
disks, even in LSB galaxies with slowly-rising rotation curves.

\vspace{0.2cm}

\end{abstract}

\keywords{galaxies: general --- galaxies: evolution --- galaxies: structure 
   --- galaxies: spiral }
}

\twocolumn[\head]

\section{Introduction} \label{sec:intro}

An exponential stellar light profile over 4-6 disk scale lengths
is an almost universal observational feature of disk galaxies
\citep[e.g.,][]{freeman70,dejong96}.
Yet, the production of a stellar
exponential disk over more than 3 scale-lengths
during the early stages of galaxy formation can prove highly 
challenging \citep[e.g.,][although see also Contardo, Steinmetz \&
Fritze-von Alvensleben 1998]{dalc97,vdb01,fer01}.
Ever since their inception \citep{lin87}, viscous evolution models
have endured as an attractive way of producing
exponential stellar disks over $\ga$4 disk scale lengths, {\it almost
independent of the initial density distribution}.  In a differentially
rotating gas disk, viscosity caused by non-circular gas motions
and turbulence transports angular momentum 
outwards as mass flows inwards \citep{pringle81,lin87}.
To prevent viscous evolution from reaching its logical endpoint
(all the mass at the origin, all the angular momentum at infinity),
the star formation (SF) timescale $t_*$ should
be within half an order of magnitude of the viscous 
timescale\footnote{This expression for $t_{\nu}$ only
applies in the case of a rotation curve which is {\it not} close to
solid body.}
$t_{\nu} \sim r^2/\nu$ 
to `freeze in' a nearly exponential stellar surface brightness
profile over many disk scale-lengths 
\citep[e.g.,][]{lin87,clarke89,yoshii89,olivier91,hellsten92,firmani96,fer01,slyz01}.

In 1D, the evolution of the gas density $\Sigma_g$ 
at a given radius $r$ is given by:
\begin{equation}
\frac{\partial\Sigma_g}{\partial t} = - \frac{1}{r} \frac{\partial}{\partial r}
	\left\{  \frac{(\partial/\partial r)[\nu \Sigma_g r^3 
	d\Omega/dr]}{d/dr (r^2 \Omega)} \right\} - \psi_*,  \label{eqn:visc}
\end{equation}
where $t$ is the time, 
$\nu$ is the viscosity, $\Omega$ is the angular velocity, and 
$\psi_*$ is the SF rate \citep{lin87}.  A brief inspection 
of Eqn.\ \ref{eqn:visc} shows a well-known result: in a galaxy
without shear (a solid-body rotation curve with $d\Omega/dr = 0$) there can be 
{\it no viscous evolution}.  

Bearing this in mind, it is interesting to note that 
recently published high-resolution ($\la$2\arcsec) \ha rotation curves for
low surface brightness (LSB) galaxies are slowly rising, and in a few cases
are nearly solid-body, within the optical extent of the galaxy
\citep[e.g.,][]{deblok01,deblok02,matthews02}.
Solid-body rotation curves have no shear, meaning that
viscous evolution would not occur.  Yet, many 
LSB galaxies have exponential surface brightness
profiles over 4 or more disk scale lengths 
\citep[e.g.,][]{mcgaugh94,deblok95,bell00}.
This could argue against a significant
role for viscous evolution in generating the exponential stellar 
disks of LSB galaxies, implying that the structure of 
(at least) LSB disks offers important insight into the initial
conditions of galactic disk formation. 

To understand if constraints on galaxy formation models can 
be gleaned from the structure of LSB galaxy disks, it is
important to quantitatively explore
this issue.  The key questions are: {\it i)} how close
to solid-body are the rotation curves of LSB galaxies? and {\it ii)} 
will the same kind of viscous evolution processes postulated
to affect Milky Way-type galaxies be able to capitalize on 
any small departures from solid-body behavior, and be able to 
generate exponential stellar disks in LSB galaxies?

To address this problem, I set out in this paper to 
estimate the possible effect of viscous
evolution in LSB galaxies with known, well-constrained
surface brightness profiles and rotation curves.
I make the central assumption 
that viscous processes in LSB galaxies and normal
Milky Way-like galaxies have the same efficiency, and that
any differences in viscous timescales are the {\it result of 
different rotation curves and gas densities only}.  
Comparing these viscous timescales for LSB
galaxies with realistic SF timescales \citep[e.g.,][]{bell00}, 
one can say whether 
the same sources of viscosity that are applicable in normal
galaxies could affect the evolution of LSB galaxies.

The plan of this paper is as follows.
In \S \ref{sec:visc}, 
a brief description of the model is given.
In \S \ref{sec:mw}, the model is calibrated to 
produce an exponential stellar light distribution for
the Milky Way.  In \S \ref{sec:lsb}, the Milky Way-calibrated
model is applied to five LSB galaxies.  I present my conclusions in 
\S \ref{sec:conc}.  As an aside, I discuss viscosity
from gas cloud collisions in Appendix \ref{app}.

\section{The Model} \label{sec:visc}

\begin{figure}[tbh]
\vspace{-0.5cm}
\hspace{-0.3cm}
\epsfxsize=\linewidth
\epsfbox{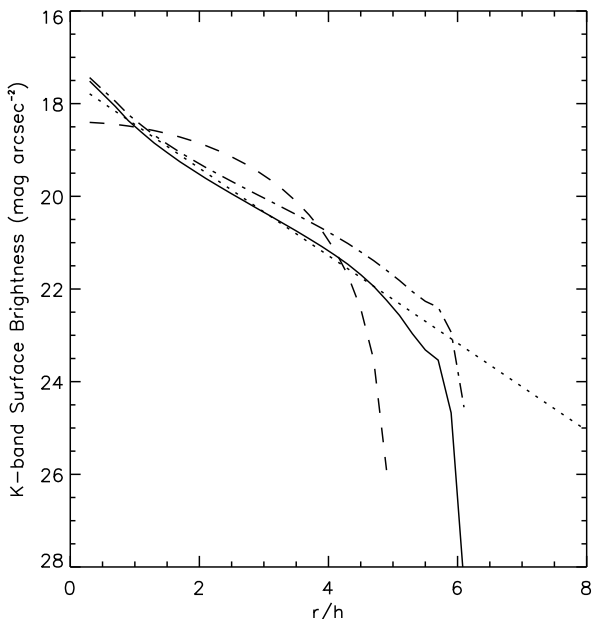}
\vspace{-0.3cm}
\caption{\label{fig:mw} 
$K$-band surface brightness profile of a Milky Way model
with a gas density-dependent SF law, which evolves subject to 
viscous evolution from gravitational instabilities.  The solid line denotes
the final $K$-band surface brightness profile, the dotted
line denotes an exponential fit to the surface brightness profile
for $r \le 6h$, the dot-dashed line shows the star$+$gas 
surface density profile assuming a
$K$-band stellar M/L of 1 in solar units, and the dashed line denotes 
the initial gas density profile, again assuming M/L$_K = 1$.  The
dynamical time SF law yields similar results.
}
\end{figure}

To investigate the role of viscous 
evolution in LSB galaxies, I modify the model of \citet{papiii} to include 
viscous evolution.  Following \citet{lin87}, I adopt a smooth initial
gas surface density distribution $\Sigma_g(r)$:
\begin{equation}
\Sigma_g(r) = \left\{ \begin{array}{ll}
   \Sigma_0 [1 + \cos(\pi r / 5 h)] & 0 \le r \le 5h \\
   0 & r > 5h,
   \end{array} \right.
\end{equation}
where $r$ is the radius, $h$ approximates the final exponential
scale-length, and 2$\Sigma_0$ is the central surface density (the final
exponential central surface density would be roughly $5 \Sigma_0$; 
see the dashed line in Fig.\ \ref{fig:mw}).
The choice of this particular initial profile is arbitrary:
\citet{yoshii89} demonstrate that exponential disks are
produced by viscous evolution from any smooth, centrally-concentrated
initial profile.  The evolution of gas density with time as
a function of radius is followed using Eqn.\ \ref{eqn:visc}.

SF is followed using either {\it i)} a gas density-dependent
SF law \citep[e.g.,][]{schmidt59} $\psi_* = k \Sigma_g^n$,
where $k$ is the rate of SF at a gas surface density
of 1 \msun\,pc$^{-2}$, and $n$ dictates how sensitively
SF rate depends on gas surface density, or {\it ii)}
a dynamical timescale-dependent 
SF law \citep[e.g.,][]{k98} $\psi_* = K \Sigma_g / \tau_{\rm dyn}$,
where $K$ is the rate of SF at a gas surface density
of 1 \msun\,pc$^{-2}$ and dynamical time of 1 Gyr, and 
$\tau_{\rm dyn} = 6.16 r({\rm kpc}) / V({\rm km\,s}^{-1})$ Gyr is
the time taken to orbit the galaxy at a distance $r$.  Both
SF laws are reasonably consistent with the SF rates and histories
of present-day spiral galaxies \citep[e.g.,][]{k98,papiii}.  
In the following, values of $k$, $n$, $K$ and the chemical 
element yield are fixed at the values
quoted in Table 1 of \citet{papiii}.

At present, sources of viscosity in galactic disks are not
well-understood.  In previous studies, the SF and viscous
timescales, $t_*$ and $t_{\nu}$, were 
directly related (where, again, $t_{\nu}$ only
applies where the rotation curve is {\it not} nearly solid body).  
The proportionality constant $\beta = t_*/t_{\nu}$ 
was constrained to be within half an order of magnitude of unity in order
to produce exponential stellar disks
over 4--6 disk scale lengths \citep[e.g.,][]{lin87,hellsten92,fer01}.
In this work, I wish to relax this constraint to a certain 
extent.  In particular, one of the things that 
I wish to assess is if $\beta \ll 1$ for LSB 
galaxies, as this would imply 
a minor role for viscous evolution in LSB galaxies.

In this paper, I explore viscosity due to gravitational instabilities.
Since I find later that viscous evolution in Milky Way-type and 
LSB galaxies is efficient with this physically-motivated prescription, 
it is sufficient
to adopt only one viscosity prescription for the purposes of this paper.
Following, e.g., \citet{olivier91} I also explored viscosity due
to collisions between gas clouds; however, I found that cloud-cloud
collisions are incapable of driving significant viscous evolution 
in any spiral galaxies (see Appendix \ref{app} for the derivation of this
result).  I therefore do not consider this prescription further.
Large-scale
gravitational instabilities give a viscosity \citep{linpringlegrav,olivier91}:
$\nu = a_{\rm grav} \pi^3 G^2 \Sigma_g^2 / 
\kappa^3$ where $a_{\rm grav}$ is a constant, $G$ is the 
gravitational constant, and $\Sigma_g$ is the gas density 
in \msun\,pc$^{-2}$.  
The epicyclic frequency $\kappa$ in Gyr$^{-1}$ is given by
$\kappa = \{ r\,\, d/dr(\Omega^2) + 4 \Omega^2 \}^{1/2}$.
I set the viscosity from gravitational instabilities to zero
if the disk is gravitationally stable, i.e., if the 
\citet{toomre} $Q > 1$, where $Q = v_s \kappa / \pi G \Sigma_g$.
The gas velocity dispersion $v_s$ is assumed to be 
6\,km\,s$^{-1}$ \citep{k89,binney}.
The constant $a_{\rm grav}$ is set to 
produce an exponential stellar disk for the Milky Way 
(essentially, 
$\beta$ for the cloud-collision and gravitational instability cases
is set to be $\sim 1$:
see \S \ref{sec:mw}).  This value of $a_{\rm grav}$ is
then left alone, and the LSB galaxy models are run to 
test if the difference in rotation curve shape and gas density
significantly affects $\beta$.  This is the central assumption 
of this analysis.

Stellar populations are modeled using the Bruzual \& Charlot
(in preparation) stellar population synthesis 
models \citep[see, e.g.,][]{liu00} 
adopting a Salpeter IMF with a reduced number of low mass stars
\citep[following][]{ml}.  I adopt the instantaneous 
recycling approximation and the closed box
approximation \citep{papiii}.
The detailed viscous
flow of metals is not tracked in this model.  These approximations
do not affect the $K$-band surface brightness or $B - R$ color
significantly, and therefore will not affect the
conclusions of this paper.  It is worth noting that use of 
the $B$-band or stellar mass surface density profiles would
also not affect the conclusions of this paper.

I solve these equations using a standard first order 
explicit scheme with 50 equally-spaced radius steps between
0.2$h$ and 10$h$.  Time steps are 4 Myr, for a galaxy age of
12 Gyr.  There are two runs per galaxy model with 
the different SF laws.
Following e.g., \citet{lin87} and \citet{fer01}, I allow mass to 
be lost from the center of the galaxy (i.e., total galaxy mass is 
not conserved); however, in practice less than 10\% of 
the total mass is lost for even large amounts of 
viscous evolution.  This does not significantly
affect my conclusions, as the galaxies with large amounts of
viscous evolution are typically quite luminous, and this `lost'
mass is easily accommodated within a bulge component.

\section{Reproducing the Milky Way} \label{sec:mw}

\begin{table*}
\caption{Parameters for galaxy models}
\label{tab}
\hspace{1.5cm}\begin{minipage}{175mm}
    \begin{tabular}{lccll}
      \hline
      \hline
Galaxy Name & $\Sigma_0$ & $h$ & $V$ & \\
 & (\msun\,pc$^{-2}$) & (kpc) & (km\,s$^{-1}$) & \\
      \hline
Milky Way & 200 & 2.82 & 220 & \\
ESO-LV 1870510 & 16 & 3.46 & $52 (1 - e^{-r[{\rm kpc}]/2.8})$ & \\
UGC 11557 & 120 & 2.77 & $110 (1 - e^{-r[{\rm kpc}]/5.1})$ & \\
F568-3 (solid-body) & 40 & 2.96 & 
	$19.95 \, r[{\rm kpc}]$ & $r \leq 3.85 {\rm kpc}$ \\
  & &  & 
	$76.8 + 9.22(r[{\rm kpc}]-3.85)^{1/2}$ & $r > 3.85 {\rm kpc}$ \\
F568-3 (1$\sigma$ curvature) & 40 & 2.96 & 
	$19.95 \, r[{\rm kpc}] + 6(1 - 0.27\{r[{\rm kpc}] - 1.923\}^2)$ 
	& $r \leq 3.85 {\rm kpc}$ \\
  & &  & 
	$76.8 + 9.22(r[{\rm kpc}]-3.85)^{1/2}$ & $r > 3.85 {\rm kpc}$ \\
F583-1 & 20 & 1.71 & $90 (1 - e^{-r[{\rm kpc}]/3.26})$ & \\
F583-4 & 16 & 2.14 & $65 (1 - e^{-r[{\rm kpc}]/1.66})$ & \\
      \hline
    \end{tabular}
\end{minipage}
\end{table*}

For the Milky Way, I adopt a flat rotation curve at all radii
with $V = 220$\,km\,s$^{-1}$ (adopting a more realistic rotation curve
in the inner parts does not significantly change the results). I choose 
a value of $h = 2.82$\,kpc, and
$\Sigma_0 = 200$\,\msun\,pc$^{-2}$, 
resulting in disk central surface densities of 
$\sim 1000$\,\msun\,pc$^{-2}$ (see Table \ref{tab} for the modeling
parameters of this and subsequent galaxy models).  These values
of $V$, $h$ and the disk central surface density are in 
reasonable accord with observations \citep[e.g.,][]{binney2}.  I assume
that the gas disk was assembled instantaneously 12 Gyr ago
(assuming more realistic infall histories for any of the models
in this paper does not significantly change the results).
A choice of $a_{\rm grav} = 0.02$ for the gravitational 
instability model results in a profile which is exponential to within
0.11 mag at $r \le 5h$ for the gas-density dependent SF law, and to within
0.14 mag for the dynamical time SF law.  In both cases, 
$B - R = 1.1 \pm 0.02$, the gas fraction is 23\% $\pm$ 3\%, and the
$K$-band central surface brightness is $17.35\pm0.15$ mag\,arcsec$^{-2}$.
These global parameters are typical 
of a galaxy of the Milky Way's luminosity and size
\citep[e.g.,][]{bdj}.  Variations of $a_{\rm grav}$ 
by factors of 2--3 do not significantly affect the 
surface brightness profiles of the model Milky Way-type galaxies:
they are still exponential to $\la$20\% within 4 disk scale lengths.

There are two interesting points which deserve further discussion.
Firstly, my value of $a_{\rm grav}$ 
results in much slower viscous evolution than estimated by \citet{olivier91}.
\citet{olivier91} estimate short viscous timescales $\la $0.2\,Gyr for
gas at $\sim$8\,kpc from the galactic center.  In order to 
keep $\beta \sim 1$, it is necessary to have SF timescales 
$\sim 0.2$\,Gyr.  These short timescales violate measured
SF timescales in large spiral galaxies in general 
\citep[e.g.,][]{k98,bdj}, and in the Milky Way in particular
\citep[$\ga 10$\,Gyr in the solar cylinder;][]{rochapinto00}.
In addition, their 
viscous flows would have velocities $\sim 5$\,km\,s$^{-1}$, violating
the $\la 1$\,km\,s$^{-1}$ constraint derived 
from radial metallicity distributions
\citep{lacey85,clarke89}.  In stark contrast, 
the `low' viscous efficiencies that I adopt result in viscous and SF timescales
of $\sim 10$\,Gyr at the solar cylinder, and gaseous flow velocities
of $\sim 0.1$\,km\,s$^{-1}$: well within observational constraints.
Secondly, the model surface brightness
profile is exponential to $\sim$10\% despite the ratio of the 
SF and viscous timescales, $\beta = t_*/t_{\nu}$, 
varying with radius by nearly an order of magnitude.
This strengthens the conclusion of \citet{hellsten92},
who stated that variations in $\beta$ by a factor of two 
with radius resulted in an exponential profile: in fact,
an order of magnitude variation in $\beta$ with radius
is acceptable in some situations.  

\section{Are LSB Galaxies Affected by Viscous Evolution?} \label{sec:lsb}

The stage is now set to address the role of viscous evolution
in LSB galaxies.  All of the ingredients are in place:
{\it i)} accurate observations of rotation curves  
for real LSB galaxies with well-constrained
exponential profiles (required for
Equation \ref{eqn:visc}), {\it ii)} 
knowledge about SF timescales in LSB galaxies
and plausible SF laws \citep{papiii}, and {\it iii)}
estimates of the efficiency of gravitational instability viscosity 
in Milky Way-type galaxies from \S \ref{sec:mw}.  The central
assumption of this comparison is that the efficiency of 
viscous processes in LSB galaxies is the same as for
Milky Way-type galaxies (i.e.,\ $a_{\rm grav}$ is
the same for all galaxies).  With this assumption, I am 
postulating that the physical processes in LSB and Milky Way-type
galaxies are identical, and differences in behavior result from
differences in {\it rotation curve and gas density only}.  
I will present the analysis for 
one particular LSB galaxy, ESO-LV 1870510, before 
extending the analysis to a further four suitable LSB galaxies.

\subsection{ESO-LV 1870510}

\begin{figure}[tbh]
\hspace{-0.3cm}
\epsfxsize=\linewidth
\epsfbox{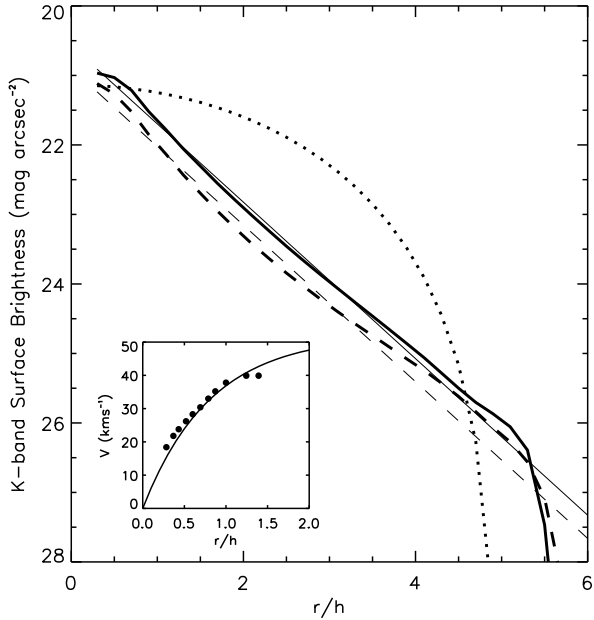}
\vspace{-0.3cm}
\caption{\label{fig:eso} 
$K$-band surface brightness profiles of two ESO-LV 1870510 models
which evolve subject to gravitational instability viscosity.
The gas density-dependent SF law is shown as a thick solid
line and the dynamical time SF law is shown by a thick dashed line.
The exponential fits to both models are also shown 
(thin solid and dashed lines).
The dotted line denotes the initial gas density profile,
assuming M/L$_K = 1$.  Inset is the smoothed rotation 
curve of \protect\citet{deblok01} and my fit to the rotation 
curve (solid line).  
}
\end{figure}

ESO-LV 1870510, at a distance of 29 Mpc \citep{bell00}, has a 
slowly-rising, almost solid-body, \hans-derived rotation curve from 
\citet[see their Fig.\ 1 and the inset panel
in Fig.\ \ref{fig:eso}]{deblok01}.  \citet{bell00} present
optical and near-IR photometry (their Fig.\ 1), which demonstrates that the 
galaxy is exponential to within 0.1 mag over 4 disk 
scale lengths.  ESO-LV 1870510 has an intrinsic
$B$-band central surface brightness of 23.75$\pm$0.2 mag\,arcsec$^{-2}$,
implying nearly a factor of 10 lower surface density than the canonical
21.65 mag\,arcsec$^{-2}$ of \citet{freeman70}.  
For the model, I adopt
the following smooth approximation to de Blok's smoothed rotation
curve: $V = 52 (1 - e^{-r[{\rm kpc}]/2.8})$\,km\,s$^{-1}$
(see inset to Fig.\ \ref{fig:eso}), a value of
$\Sigma_0 = 16$\,\msun\,pc$^{-2}$ and $h = 3.46$\,kpc (see also Table 
\ref{tab}).

Fig.\ \ref{fig:eso} shows the resulting $K$-band surface 
brightness profiles of the density-dependent and dynamical
time SF law LSB model galaxies.
Clearly, the viscous redistribution and SF timescales 
turn out to be roughly equal, and exponential surface brightness
profiles are produced.  The stellar disks are
exponential to within 0.07 mag (density-dependent SF law)
or 0.13 mag (dynamical time SF law)  within
$r < 4h$.  The model $B - R$ colors match the observations to $\la 0.1$ mag, 
and the scale lengths, magnitudes and 
central surface brightnesses match the observations to 20\%.
ESO-LV 1870510 lacks literature \hi data, however, 
the model gas fraction of $\sim$60\% is 
typical of LSB galaxies \citep[e.g.,][]{bdj}.

\subsubsection{Why Can Viscous Evolution Affect LSB Galaxies?}

It is interesting to consider why viscous evolution can 
work with similar efficiency in LSB galaxies and Milky Way-type
galaxies.  After all, the rotation curves of LSB galaxies rise
slowly (see, e.g., the inset panel in Fig.\ \ref{fig:eso}), implying
that viscous evolution could be much less efficient.  
Using that $t_{\nu} \sim r^2/\nu$, $t_{\nu} \propto r^2 \kappa^3/\Sigma_g^2$. 
Fig.\ 3 of \citet{dejong00} shows that Milky Way-type and LSB galaxies
have similar scale lengths, therefore characteristic
values of $r$ are similar for LSB and Milky Way-type galaxies
(see also Table \ref{tab}).
The gas density $\Sigma_g$ is 
a factor of $\sim 10$ lower in LSB galaxies, which drives
up $t_{\nu}$.  However, $\kappa \sim \Omega \propto V/r$ \citep{binney} is 
substantially lower in LSB galaxies as $r$ is similar but
$V$ is a factor of a few lower.  
Consequently $\kappa^3/\Sigma_g^2$,
therefore $t_{\nu}$, are similar for Milky Way-type and LSB galaxies.

Considering the rest of Equation 
\ref{eqn:visc}, we see that 
the top term in the equation is reduced
for a LSB galaxy by a factor of 100: a factor of 10 from the $\Sigma_g$
term and a factor of 10 from $d\Omega/dr$.  The bottom term
is reduced by a factor of $\sim$5, simply due to the 
more solid-body rotation curve shape.  Therefore the ratio, 
and the derivative of the ratio, is 
roughly a factor of 20 lower for a LSB galaxy.  However, the gas densities
are roughly a factor of 10 lower, so $\{\partial \Sigma_g / \partial t\} / 
\Sigma_g$ is only reduced by a factor of 2 or so: viscous evolution
from gravitational instability will be nearly as effective in 
LSB galaxies as it would be in large spiral galaxies!  Coupled with 
the somewhat longer $t_*$ in LSB galaxies, it is clear that
$\beta = t_*/t_{\nu}$ will remain of order unity, easily producing an 
exponential disk.

\subsubsection{Gravitational Instability and LSB Galaxies}

I have demonstrated that viscous evolution through
gravitational instability can be just as effective in LSB galaxies
as in Milky Way-type galaxies.  However, this seems to contradict
at some level the argument that LSB galaxies are rather more
likely the be gravitationally stable than their higher surface
brightness counterparts.
If LSB galaxies were gravitationally stable,
no viscous evolution (at least through gravitational instabilities) 
would occur.
\citet{vdh93} estimate the Toomre $Q$ parameter,
finding $1 \la Q \la 3$.  This can be compared to $0.5 \la Q \la 2$ for 
most luminous spiral galaxies in their inner parts \citep{k89}.
However, there are reasons to believe that \citet{vdh93} could
significantly overestimate $Q$.  Toomre's $Q \propto 1/\Sigma_g$,
and poor resolution \hi observations will tend to underestimate
$\Sigma_g$.  In addition, it is unclear how much molecular
hydrogen there is in LSB galaxies, due primarily to the 
uncertainty in the CO to H$_2$ ratio $X$ 
\citep[see, e.g.,][]{mihos,boselli}: a 50\% 
increase in $\Sigma_g$ caused by molecular hydrogen in 
the inner few scale lengths is not unreasonable given that $X$ could
be a factor of 10 higher than galactic, and given
the level of current CO detections of LSB galaxies 
\citep{matthews01}.  Both of these
effects would tend to drive down $Q$, making LSB galaxies less
stable.  

However, the model galaxy has gas densities which are 
somewhat larger than typical observed densities in LSB galaxies
(10 \msun\,pc$^{-2}$ at the model half-light radius, compared to 
$\sim$5 \msun\,pc$^{-2}$ for observed LSB galaxies).
This problem can be alleviated by allowing the model galaxy 
to be built up by a lengthy gas infall.  Assuming an `inside-out'
formation scenario with an $e$-folding timescale of 2.5 Gyr at the 
center, and 10 Gyr at $\sim 4 h$ \citep[the infall case of][]{papiii}, 
the gas density of the model
galaxy is reduced by a factor of two, bringing it much closer
to typical, observed gas densities in LSB galaxies.  However, viscous evolution
still occurs, producing a stellar exponential disk: 
the viscous evolution basically operates at 
a level which keeps the LSB galaxy on the edge of gravitational
instability throughout its life.
Given the considerable observational and
modeling uncertainties, it is not unreasonably conservative to argue that
LSB galaxies are at the edge of gravitational instability
for most of their lives \citep[note that][also argue for locally
unstable LSB disks]{mihos97}, and therefore could plausibly undergo significant
viscous evolution.

\subsection{Other LSB galaxies} \label{subsec:others}

\begin{figure}[tbh]
\hspace{-0.3cm}
\epsfxsize=\linewidth
\epsfbox{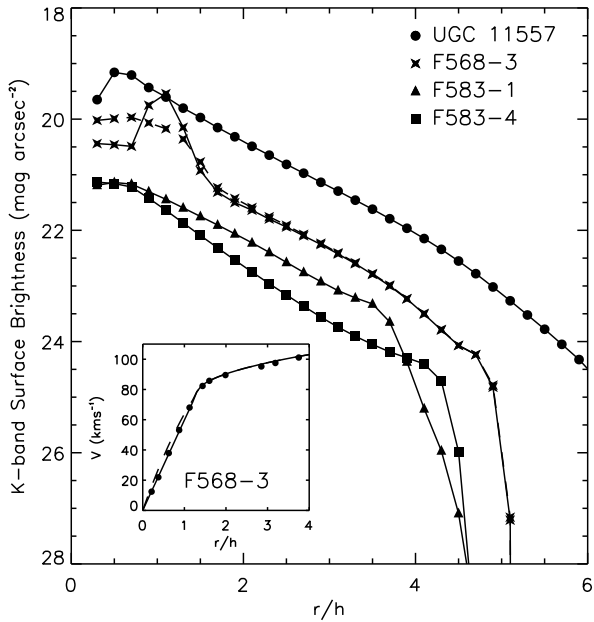}
\vspace{-0.3cm}
\caption{\label{fig:other} 
$K$-band surface brightness profiles of the models of four LSB galaxies from
the sample of \protect\citet{deblok01}.  The gas density-dependent
models, which evolve due to viscosity from gravitational instabilities, 
are shown for 
all galaxies.  F568-3 has two models shown: one with a solid-body
inner rotation curve (solid line) and one with curvature which is 
consistent with the 1$\sigma$ errors in the rotation curve (dashed line).
With the exception of the solid-body case of F568-3, all surface 
brightness profiles do not deviate from exponential by more
than $\sim$0.1 mag.  The use of the dynamical time SF law  
would not affect the results significantly.
}
\end{figure}

I have demonstrated that it is possible for the LSB galaxy ESO-LV
1870510 to undergo significant viscous evolution if it is gravitationally
unstable for at least part of its life.  In this subsection,
I repeat the analysis with a further four galaxies with
accurate, modelled
rotation curves \citep{deblok01} and exponential profiles to good accuracy
over at least three disk scale lengths \citep{deblok96,swaters99,bell00}: UGC
11557, F568-3, F583-1 and F583-4 (see Table \ref{tab} for modeling parameters
and Fig.\ \ref{fig:other} for the results).

In all cases the gas disks
can be unstable, and this leads to significant viscous evolution with
either the gas density-dependent or dynamical time-dependent SF law.
The cloud collision viscosity model gives very little viscous evolution.
None of the rotation curves are solid body: most rise slowly out to
$\sim$2$h$, and turn over at larger radii.  This is common
amongst LSB galaxies \citep[e.g.,][]{deblok01,matthews02}.  The only
exception to this is F568-3, which has a rotation curve within
$\sim$1.2$h$ which can be modelled as solid body.  This leads to a `ring'
feature in the viscous model surface brightness profile, where material
piles up at $r \sim h$ (Fig.\ \ref{fig:other}, solid line with stars).
If the rotation curve is allowed to `curve'
slightly at the observational $1\sigma$ level, the final surface brightness
profile smooths considerably, and becomes exponential to within 0.1 mag
RMS (Fig.\ \ref{fig:other}, dashed line with stars).

This strengthens the conclusions considerably: it is impossible
to rule out a significant role for viscous evolution during the life of
five LSB galaxies with exponential surface brightness
profiles for which accurate rotation curves are available from the literature.

\section{Conclusions} \label{sec:conc}

Models of galaxy evolution which include viscosity due to non-circular
and turbulent gas motions have, for the last 15 years, provided
a relatively appealing way of producing accurately exponential stellar
disks in large spiral galaxies almost independent of initial 
conditions.  The primary assumption is that the viscous timescale
and star formation timescale should be approximately equal.
However, galaxies with solid-body rotation curves cannot undergo viscous
evolution.  Low surface brightness (LSB) galaxies have exponential surface
brightness profiles, yet have slowly rising, nearly
solid-body, rotation curves.  Thus, it could be that LSB
galaxies have had little redistribution of gas due to viscous
evolution during their lifetimes.  Their surface brightness profiles
would give unique insight into the initial conditions of galaxy formation.

In this paper, I have constrained the importance of viscous
evolution during the life of LSB galaxies.  I have
adopted star formation laws from the literature, and have tuned
the efficiency of viscous processes to reproduce an exponential stellar
profile in Milky Way-type galaxies.  Conservatively, I have assumed
that the physics of turbulence are the same in LSB and Milky Way-type
galaxies: any differences in the importance of viscous evolution in
the two types of galaxy are the product of differences in rotation
curve and gas density only.  I then apply this model to LSB galaxies
with observed rotation curves and surface brightness profiles.
Under the conservative and not unreasonable condition that
LSB galaxies are gravitationally unstable for at least part of their
lives, I find that viscous evolution from gravitational
instabilities is quite effective.
Viscous evolution models
still offer an attractive way of producing an exponential disk
over four or more disk scale lengths, almost independent
of initial density distribution, even for LSB galaxies
with slowly-rising rotation curves.

Does this have implications for the power of surface brightness
profiles to constrain galaxy formation theory?  It is 
impossible at present to predict the degree of viscous evolution 
that {\it will} occur, as the physical mechanisms driving
viscous evolution are so poorly constrained.  However, this paper
has shown that it {\it can} plausibly occur in both low and high surface
density disk galaxies, and \citet{yoshii89} have shown that it
would modify most intial density profiles into an exponential disk.
This reduces substantially the ability of 
observations to rule out models of galaxy formation 
based on surface brightness profiles.  Indeed, the assembly
of a disk with a reasonably centrally concentrated profile with 
roughly the right amount of angular momentum is all that is 
required of a galaxy formation model.  However, it is impossible
to reduce the central matter density of a galaxy through
viscous processes; therefore, models which overpredict the 
central mass density of galaxy disks cannot escape the need for
revision or modification \citep[see, e.g.,][]{vdb01}.

\acknowledgements

I thank Dennis Zaritsky, Erwin de Blok and St\'ephane Court\-eau 
for useful discussions during the course of this work.  
Special thanks go to the anonymous referee, whose insightful comments
led me to examine cloud-collision viscosity more closely.
I was supported by NSF grant AST-9900789.

\appendix
\section{A. Exploring cloud-cloud viscosity} \label{app}

Here, I briefly discuss cloud-cloud 
collision viscosity.  Despite its frequent use in the literature,
it has not been comprehensively derived and so I will explore it here.
Following, e.g.,
\citet{silk81} and \citet{olivier91} I assume a cloud-cloud 
collision rate $R \sim n_{\rm cloud} v_s \pi r^2_{\rm cloud}$,
where $n_{\rm cloud}$ is the number space density of clouds,
$v_s$ is the gas velocity dispersion, and 
$r_{\rm cloud}$ is the typical cloud radius.  Assuming that the cloud
properties and galaxy scale height are independent of gas surface
density, $n_{\rm cloud} = \Sigma_g / (M_{\rm cloud} h)$, where
$\Sigma_g$ is the gas surface density, $M_{\rm cloud}$ is the 
typical gas cloud mass and $h$ is the typical gas scale height.
Adopting the cloud-collision parameters from \citet{olivier91},
$R \sim 0.032 v_s \Sigma_g$\,Gyr$^{-1}$ where $\Sigma_g$ is in \msun\,pc$^{-2}$
and $v_s$ is in km\,s$^{-1}$
\citep[so far, this derivation does not significantly 
deviate from][]{olivier91}.

Following \citet{pringle81}, the kinematic viscosity takes the 
basic form $\nu \sim v_s \lambda$, where $\lambda$ is the 
effective mean free path of clouds before they collide.  In the
limit that cloud collisions are much more frequent than once
per orbit, $\lambda$ is equal to the mean free path of the clouds
which is $v_s \tau_c$, where $\tau_c = 1/R$ is the cloud-collision
timescale.  This yields a viscosity 
$\nu = 32.6 \eta_{\rm freq} v_s / \Sigma_g$\,kpc$^2$\,Gyr$^{-1}$.
In the limit of cloud collisions being much less frequent than once
per orbit, $\nu \sim v_s \lambda_r (\tau_c/\tau_{\kappa})^{-1}$
where $\lambda_r \sim v_s \tau_{\kappa}$ is the radial excursion 
during one epicycle, and $\tau_{\kappa} = 2\pi / \kappa$ is the
epicycle timescale \citep{pringle81}.  In this case, 
$\nu \sim v_s^2  \tau_{\kappa}^2 / \tau_c = 1.31 \eta_{\rm infreq} v_s^3 \Sigma_g / \kappa^2$\,kpc$^2$\,Gyr$^{-1}$, 
where $v_s$ is in km\,s$^{-1}$, $\Sigma_g$ is in \msun\,pc$^{-2}$,
and $\kappa$ is in Gyr$^{-1}$.  
Which limit is adopted depends on $\Sigma_g$ (assuming that $v_s$ is constant
between different galaxy types).  For gas densities
$\ga$25 \msun\,pc$^{-2}$, the frequent collision case is more appropriate
($\tau_c \la 200$ Myr), and for gas densities $\la$25 \msun\,pc$^{-2}$
the infrequent case is more appropriate ($\tau_c \ga 200$ Myr), for 
a constant $v_s \sim 6$ km\,s$^{-1}$.

The crucial issue now is to estimate the efficiencies 
$\eta_{\rm freq}$ and $\eta_{\rm infreq}$ of the frequent and 
infrequent cloud-cloud collision cases.  This is done by estimating
the energy loss per cloud-cloud collision.  The specific kinetic energy
of a cloud is $\sim v_{\rm rot}^2$, where $v_{\rm rot}$ is the 
rotation velocity.  However, clouds collide with speeds much less than
than $v_{\rm rot}$ (because they are almost always in very similar, nearly
circular orbits).  In this case, the maximum possible specific
energy loss is related to their relative speeds $\sim v_{\rm relative}^2$.
Therefore, the efficiency of energy loss per collision is
$\sim (v_{\rm relative}/v_{\rm rot})^2$.  
The relative velocity $v_{\rm relative}$ can 
be approximated by the difference in velocity
caused by the radial excursion of the cloud.  Using that 
$v_{\rm relative} = v_{\rm rot} \Delta r / r$, where $r$ is the
radius, and substituting $v_{\rm rot} = r \Omega$ ($\Omega$ is 
the angular frequency) and $\Delta r \sim \lambda$, 
$v_{\rm relative} \sim \Omega \lambda$.
For the frequent collision case $\lambda \sim \tau_c v_s$, therefore
$v_{\rm relative} \sim \Omega / 0.032 \Sigma_g$ (where 
$\Omega$ is in Gyr$^{-1}$ and 
$\Sigma_g$ is in \msun\,pc$^{-2}$).  Thus, 
$\eta_{\rm freq} \sim 977/(r^2 \Sigma^2)$, where $r$ is in kpc.
For the infrequent
collisions case $\lambda$ is the maximum epicyclic 
excursion, which is roughly $v_s$ multiplied by the time for
1/4 epicycle $\sim v_s / \kappa$.  Using that $\kappa/\Omega \sim 1$, 
$v_{\rm relative} \sim v_s$ for the infrequent case, where $v_s$ is 
in km\,s$^{-1}$.  
Therefore, $\eta_{\rm infreq} \sim v_s^2 / (r^2 \Omega^2)$.

What are reasonable estimates of the timescales of viscous
evolution in the frequent (Milky Way-type galaxy) and infrequent 
(LSB galaxy) cases?  Using $r = 7.5$\,kpc, $v_{\rm rot} = 220$\,km\,s$^{-1}$,
$v_s = 6$\,km\,s$^{-1}$, and $\Sigma_g = 50$\,\msun\,pc$^{-2}$
for a Milky Way-type galaxy, I find that
$v_{\rm relative} \sim 20$\,km\,s$^{-1}$ and
$t_{\nu} \sim r^2/\nu \sim 2000$\,Gyr.
Similarly, adopting $\Sigma_g = 10$\,\msun\,pc$^{-2}$ and
$v_{\rm rot} = 100$\,km\,s$^{-1}$ for the LSB galaxy infrequent case, 
I find $t_{\nu} \sim 1000$\,Gyr.  Thus, it is fair to conclude that 
{\it viscosity from cloud-cloud collisions will be 
unimportant for all spiral galaxies}.  This is because the low
efficiency of energy loss in cloud-cloud collisions 
$(v_{\rm relative}/v_{\rm rot})^2 \ll 0.01$, which more than 
outweighs the fact that cloud collisions tend to happen $\ga 1$
time per orbit for most spiral galaxies.  Of course, if some
mechanism `whips up' cloud collision velocities (such as a galaxy
interaction), it may be possible to evolve significantly from
cloud-cloud viscosity, but this mechanism should not play 
a significant r\^ole in the evolution of spiral galaxies
in their normal, quiescent mode.

\end{document}